\documentclass[12pt]{article}
\pdfoutput=1
\usepackage{xcolor}
\usepackage{hyperref}
\usepackage{subcaption}
\usepackage[utf8]{inputenc}

\usepackage{setspace}
\usepackage{amsmath, amssymb, amsthm, float, graphicx}
\numberwithin{equation}{section}

\hypersetup{colorlinks=false, linkcolor=blue, citecolor=red}

\def\beq{\begin{eqnarray}}\def\eeq{\end{eqnarray}}
\def\be{\begin{equation}}\def\ee{\end{equation}}
\def\g{\gamma}
\def\r{\rho}

\def\m{\mu}

\def\a{\alpha}

\def\k{\kappa}
\def\b{\beta}
\def\d{\delta}

\def\D{\Delta}
\def\G{\Gamma}

\def\pd{\partial}
\def\tq{\tilde{q}}
\def\ta{\tau}
\def\bz{\bar{z}}
\def\la{\langle}
\def\ra{\rangle}

\def\mo{{\mathcal{O}}}

\def\G{\Gamma}
\def\mc{{\mathcal{C}}}
\def\bh{\bar{h}}

\begin{document}
\title{\bf Universal anomalous dimensions\\ at large spin and large twist }
\date{}

\author{Apratim Kaviraj\footnote{apratim@cts.iisc.ernet.in}, ~Kallol Sen\footnote{kallol@cts.iisc.ernet.in} ~and Aninda Sinha\footnote{asinha@cts.iisc.ernet.in}\\ ~~~~\\
\it Centre for High Energy Physics,
\it Indian Institute of Science,\\ \it C.V. Raman Avenue, Bangalore 560012, India. \\}
\maketitle

\abstract{In this paper we consider anomalous dimensions of double trace operators at large spin ($\ell$) and large twist ($\tau$) in CFTs in arbitrary dimensions ($d\geq 3$). Using analytic conformal bootstrap methods, we show that the anomalous dimensions are universal in the limit $\ell\gg \tau\gg 1$. In the course of the derivation, we extract an approximate closed form expression for the conformal blocks arising in the four point function of identical scalars in any dimension. We compare our results with two different calculations in holography and find perfect agreement. }

\vskip 1cm

\tableofcontents

\onehalfspacing

\section{Introduction}

With the recent resurgence in the applications \cite{Rattazzi, Polchinski, Hogervorst, SCFTs} of the conformal bootstrap program in higher dimensions, it is of interest to ask which of these results are universal and will hold for any conformal field theory (under some minimal set of assumptions). Among other things, such results may prove useful tests for the burgeoning set of numerical tools (see e.g. \cite{Showk}) being used in the program. In this paper we will prove one such result in any $d$-dimensional CFT with $d\geq 3$. 

We will consider CFTs in dimensions greater than two. We will assume that there is a scalar operator of dimension $\Delta_\phi$ and a minimal twist ($\tau=d-2$) stress energy tensor. Then 
using the arguments of \cite{anboot, Komargodski}, one finds that there is an infinite sequence of large spin operators of twists $\tau=2\Delta_\phi+2n$ where $n\geq 0$ is an integer. By assuming that there is a twist gap separating these operators from other operators in the spectrum, we can go onto setting up the bootstrap equations which will enable us to extract the anomalous dimension $\gamma(n,\ell)$ of these operators \cite{anboot, Komargodski, kss}. In \cite{kss}, we derived $\gamma(n,\ell)$ for 4d-CFTs satisfying the above conditions in the large spin limit. We found that the $n\gg 1$ result was universal in the sense that it only depended on $n,\ell$ and $c_T$ the coefficient appearing in the two point function of stress tensors, with no dependence on $\Delta_\phi$. We were able to show an exact agreement with the Eikonal limit calculation by Cornalba et al \cite{Cornalba1,Cornalba2,Cornalba3}. In this paper we will extend this analysis to arbitrary dimensions greater than two.

One of the key results which will enable us to perform this calculation is the derivation of a closed form expression for the conformal blocks in arbitrary dimensions in a certain approximation. This was already initiated in \cite{anboot, Komargodski} and we will take this to the logical conclusion needed to extract the anomalous dimensions. In particular, we will derive an expression that solves a recursion relation (see eq.(70) appendix A of \cite{anboot} which follows from \cite{Dolan}) relating the blocks in $d$ dimensions to the blocks in $d-2$ dimensions.  In the large spin limit, the blocks simplify and approximately factorize. This is what allows us to perform analytic calculations. 

We find that the anomalous dimensions $\gamma(n,\ell)$ in the limit $\ell\gg n\gg 1$ take on the form
\begin{align}
\gamma(n,\ell)=&-\frac{8(d+1)}{c_T (d-1)^2}\frac{\G(d)^2}{\G(\frac{d}{2})^4}\frac{n^d}{\ell^{d-2}}\,.
\end{align}
Here $c_T$ is the coefficient appearing in the two point function of stress tensors. Since $\ell\gg 1$ we do not need $c_T$ to be large to derive this result. However, if $c_T$ is large we can identify the operators as double trace operators. One of the main motivations for looking into this question was an interesting observation made in \cite{Camanho:2014apa} which relates the sign of the anomalous dimensions\footnote{see \cite{suzuki} for a recent work on the sign of anomalous dimensions in $\mathcal{N}=4$ Yang-Mills in perturbation theory.} of double trace operators with Shapiro time delay suggesting an interesting link between unitarity of the boundary theory and causality in the bulk theory. We found in \cite{kss}, that $\gamma(1,\ell)$ could be positive if $\Delta_\phi$ violated the unitarity bound.

We will show that the bootstrap result is in exact agreement with the holographic calculation performed in \cite{Cornalba1}. In this calculation, one needs to assume both large spin and large twist. Moreover, why the sign of the anomalous dimension is negative as well as why the result for the  anomalous dimension in this limit is independent of $\alpha'$ corrections are somewhat obscure. We will turn to another calculation in holography, proposed in \cite{kfw} which has three advantages: (a) one can consider $\ell \gg 1$ but $n$ not necessarily large (b) it makes it somewhat more transparent why $\alpha'$ corrections do not contribute to the leading order result except through $c_T$ and (c) it relates the negative sign of the anomalous dimension to the positive sign of the AdS Schwarzschild black hole mass. We will extend the results of \cite{kfw} who considered $n=0$ to finite $n$. 
 
Our paper is organized as follows. In section 2, we write down a closed form expression in a certain approximation for conformal blocks \cite{Dolan, Dolan2,Dolan3} in general dimensions. In section 3, we set up the calculation of anomalous dimensions (for double trace operators) in general dimensions $\geq 3$ using analytic bootstrap methods. In section 4, we perform the sums needed in the limit $\ell\gg n\gg 1$ and derive the universal result eq.(1). In section 5, we turn to holographic calculations of the same results. We conclude with a brief discussion of open problems in section 6. Appendix A shows the exact $n$ dependence  in $d=6$ extending the $d=4$ result \cite{kss}.

\section{Approximate conformal blocks in general $d$}
We start with the bootstrap equation used in \cite{ anboot, kss} \footnote{The factor of $\frac{1}{4}$ on the \textit{lhs} is to match with the conventions of \cite{kss}.},
\be\label{bssl}
1+\frac{1}{4}\sum_{\ell_m} P_m u^{\frac{\ta_m}{2}}f_{\ta_m,\ell_m}(0,v)+O(u^{\frac{\ta_m}{2}+1})= \bigg(\frac{u}{v}\bigg)^{\D_\phi}\sum_{\ta,\ell}P_{\ta,\ell}g^{(d)}_{\ta,\ell}(v,u)\,.
\ee
For general $d$ dimensional CFT, the minimal twist $\ta_m=d-2$. This is the twist for the stress tensor which we will assume to be in the spectrum. The function $f_{\ta_m,\ell_m}(0,v)$ is of the form,
\be
f_{\ta_m,\ell_m}(0,v)=(1-v)^{\ell_m}\ {}_2F_1\bigg(\frac{\ta_m}{2}+\ell_m,\frac{\ta_m}{2}+\ell_m,\ta_m+2\ell_m,1-v\bigg)\,,
\ee
On the \textit{rhs} of \eqref{bssl}, $g^{(d)}_{\ta,\ell}(v,u)$ denote the conformal blocks in the crossed channel. In the limit of the large spin, $g^{(d)}_{\ta,\ell}(v,u)$ undergo significant simplification as given in appendix A of \cite{anboot}. For any general $d$, the function $g^{(d)}_{\ta,\ell}(v,u)$ can be written as,
\be
g^{(d)}_{\ta,\ell}(v,u)=k_{2\ell+\ta}(1-u) F^{(d)}(\ta,v)+O(e^{-2\ell\sqrt{v}})\,,
\ee  
where subleading terms are exponentially suppressed at large $\ell$ and,
\be
k_\b(x)=x^{\b/2}\ {}_2F_1\bigg(\frac{\b}{2},\frac{\b}{2},\b,x\bigg)\,.
\ee
In the limit of large $\ell$ and fixed $\ta$ and for $u\rightarrow0$, following appendix A of \cite{anboot},
\be\label{fac}
g^{(d)}_{\ta,\ell}(v,u)=k_{2\ell}(1-u)v^{\ta/2}F^{(d)}(\ta,v)+O(1/\sqrt{\ell},\sqrt{u})\,.
\ee 
Thus to the leading order we need to find the functions $F^{(d)}(\ta,v)$ to complete the derivation of the factorizaion ansatz given in \eqref{fac}. 

To derive a form of the function $F^{(d)}(\ta,v)$, we start by writing down the recursion relation relating the conformal block in $d$ dimension to the ones in $d-2$ dimensions (see \cite{Dolan, anboot}),
\begin{align}\label{rec0}
\begin{split}
\bigg(\frac{\bz-z}{(1-z)(1-\bz)}\bigg)^2 g^{(d)}_{\D,\ell}(v,u)=&g^{(d-2)}_{\D-2,\ell+2}(v,u)-\frac{4(\ell-2)(d+\ell-3)}{(d+2\ell-4)(d+2\ell-2)}g^{(d-2)}_{\D-2,\ell}(v,u)\\
&-\frac{4(d-\D-3)(d-\D-2)}{(d-2\D-2)(d-2\D)}\bigg[\frac{(\D+\ell)^2}{16(\D+\ell-1)(\D+\ell+1)}g^{(d-2)}_{\D,\ell+2}(v,u)\\
&-\frac{(d+\ell-4)(d+\ell-3)(d+\ell-\D-2)^2}{4(d+2\ell-4)(d+2\ell-2)(d+\ell-\D-3)(d+\ell-\D-1)}g^{(d-2)}_{\D,\ell}(v,u)\bigg]\,,
\end{split}
\end{align}
where $u=z\bz$ and $v=(1-z)(1-\bz)$.
In the limit when $\ell\rightarrow\infty$ at fixed $\ta=\D-\ell$, \eqref{rec0} becomes,
\begin{align}\label{rec}
\begin{split}
\bigg(\frac{\bz-z}{(1-z)(1-\bz)}\bigg)^2 g^{(d)}_{\ta,\ell}(v,u)\overset{\ell\gg1}{=}g^{(d-2)}_{\ta-4,\ell+2}(v,u)-&g^{(d-2)}_{\ta-2,\ell}(v,u)-\frac{1}{16}g^{(d-2)}_{\ta-2,\ell+2}(v,u)\\
&+\frac{(d-\ta-2)^2}{16(d-\ta-3)(d-\ta-1)}g^{(d-2)}_{\ta,\ell}(v,u)+O\bigg(\frac{1}{\ell}\bigg)\,.
\end{split}
\end{align}
Furthermore for $z\rightarrow0$ and $z\ll\bz=1-v+O(z)<1$ we can write the \textit{lhs} of \eqref{rec} as,
\be
\bigg(\frac{\bz-z}{(1-z)(1-\bz)}\bigg)^2 g^{(d)}_{\ta,\ell}(v,u)\overset{u\rightarrow0}{=}\left[\bigg(\frac{1-v}{v}\bigg)^2+O(u)\right]g^{(d)}_{\ta,\ell}(v,u)\,.
\ee
Finally putting in the factorization form in \eqref{fac}, we get,
\begin{align}\label{Fd}
(1-v)^2F^{(d)}(\ta,v)=16 F^{(d-2)}(\ta-4,v)-2vF^{(d-2)}(\ta-2,v)+\frac{(d-\ta-2)^2}{16(d-\ta-3)(d-\ta-1)}v^2F^{(d-2)}(\ta,v)\nonumber\\+O(1/\sqrt{\ell},\sqrt{u})\,.
\end{align}
We find $F^{(d)}(\ta,v)$ to be \footnote{ We guessed this form of the solution by looking at the explicit forms of $F^{(d)}(\ta,v)$ for $d=2,4$.},
\be\label{Fd1}
F^{(d)}(\ta,v)=\frac{2^\ta}{(1-v)^{\frac{d-2}{2}}}\ {}_2F_1\bigg(\frac{\ta-d+2}{2},\frac{\ta-d+2}{2},\ta-d+2,v\bigg)\,.
\ee
To see whether \eqref{Fd1} satisfies the recursion relation in \eqref{Fd}, we plug in \eqref{Fd1} in \eqref{Fd} and expand both sides in powers of $v$. Then the \textit{lhs} is,
\begin{align}\label{reclhs}
&(1-v)^2F^{(d+2)}(\ta,v)=\nonumber\\&\sum_{k=0}^\infty v^k\ 2^{-2+\tau }\text{  }\frac{\left(d^2-4 k-2 d (-2+\tau )+(-2+\tau )^2\right) \Gamma ^2\left(-1-\frac{d}{2}+k+\frac{\tau }{2}\right) \Gamma (-d+\tau )}{(1-v)^{\frac{d-2}{2}}\Gamma ^2(1+k) \Gamma ^2\left(\frac{-d+\tau }{2} \right) \Gamma (-d+k+\tau )}\,.
\end{align}
The \textit{rhs} under power series expansion gives,
\begin{align}\label{Fd2}
16 F^{(d)}&(\ta-4,v)-2vF^{(d)}(\ta-2,v)+\frac{(d-\ta)^2}{16(d-\ta-1)(d-\ta+1)}v^2F^{(d)}(\ta,v)\nonumber\\&=\sum _{k=0}^{\infty }\frac{v^k}{(1-v)^{\frac{d-2}{2}}} \left(\frac{2^{-4+\tau }(d-\tau )^2 \Gamma ^2\left(-1-\frac{d}{2}+k+\frac{\tau }{2}\right) \Gamma (2-d+\tau )}{\left(-1+d^2-2 d \tau +\tau ^2\right) (-2+k)! \Gamma ^2\left(\frac{2-d+\tau }{2}\right) \Gamma (-d+k+\tau )}\right.\nonumber\\&\left.+\frac{2^{\tau } \Gamma (-2-d+\tau ) \Gamma ^2\left(k-\frac{2+d-\tau }{2}\right)}{k! \Gamma ^2\left(\frac{-2-d+\tau }{2}\right) \Gamma (-2-d+k+\tau )}-\frac{2^{\tau -1} \Gamma (-d+\tau ) \Gamma ^2\left(-1+k-\frac{d-\tau }{2}\right)}{(-1+k)! \Gamma^2 \left(\frac{-d+\tau }{2}\right) \Gamma (-1-d+k+\tau )}\right)\,.
\end{align}
Using properties of gamma functions the above series simplifies to the one given in \eqref{reclhs}\footnote{ We have checked in Mathematica that the expression in \eqref{Fd2} after FullSimplify matches with \eqref{reclhs}} . Since the two series are the same, the expression \eqref{Fd1} is the solution to the recursion relation \eqref{Fd} for any $d$\footnote{One can show that using \eqref{Fd1} in \eqref{bssl} reproduces the exact leading term 1, on the $lhs$, by cancelling all the subleading powers of $v$. This shows that \eqref{Fd1} is indeed the correct form of $F^{(d)}(\tau,v)$ for general $d$.}. With the knowledge of $F^{(d)}(\ta,v)$ for general $d$, we can now do the same analysis for the anomalous dimensions in general $d$, following what was done in \cite{kss} for $d=4$.  

\section{Anomalous dimensions for general $d$}\label{anomdimgen}

$F^{(d)}(\ta,v)$ can be written as,
\be\label{ddef}
F^{(d)}(\ta,v)=\frac{2^\ta}{(1-v)^{\frac{d-2}{2}}}\sum_{k=0}^\infty d_{\ta,k}v^k, \ \text{where},\ d_{\ta,k}=\frac{((\ta-d)/2+1)_k{}^2}{(\ta-d+2)_k k!}\,,
\ee
where $(a)_b=\G(a+b)/\G(a)$. The MFT coefficients for general $d$, after the large $\ell$ expansion takes the form,
\be
P_{2\D_\phi+2n,\ell}\overset{\ell\gg1}{\approx}\frac{\sqrt{\pi}}{4^\ell}\tq_{\D_\phi,n}\ell^{2\D_\phi-3/2}\,.
\ee
where,
\be
\tilde{q}_{\D_\phi,n}=2^{3-\ta}\frac{\G(\D_\phi+n-d/2+1)^2\G(2\D_\phi+n-d+1)}{n!\G(\D_\phi)^2\G(\D_\phi-d/2+1)^2\G(2\D_\phi+2n-d+1)}\,.
\ee
The \textit{rhs} of \eqref{bssl} can be written as,
\be
\sum_{\ta,\ell} P_{\ta,\ell}^{MFT} v^{\frac{\ta}{2}-\D_\phi}z^{\D_\phi}(1-v)^{\D_\phi}k_{2\ell}(1-z)F^{(d)}(\ta,v)\,.
\ee
The twists are given by $\ta=2\D_\phi+2n+\g(n,\ell)$. Thus in the large spin limit, to the first order in $\g(n,\ell)$ we get,
\be
\sum_{n,\ell} P_{2\D_\phi+2n,\ell}^{MFT}\bigg[\frac{\g(n,\ell)}{2}\log v\bigg] v^n \frac{\ell^{\frac{1}{2}}4^\ell}{\sqrt{\pi}}K_0(2\ell\sqrt{z}) z^{\D_\phi}(1-v)^{\D_\phi}F^{(d)}(2\D_\phi+2n,v)\,,
\ee
where $K_0(2\ell\sqrt{z})$ is the modified Bessel function. Moreover for large $\ell$, the anomalous dimensions behaves with $\ell$ like  (see \cite{anboot, kss}), \be
\g(n,\ell)=\frac{\g_n}{\ell^{\ta_m}}\,,
\ee
where $\ta_m$ is the minimal twist. With this we can convert the sum over $\ell$ into an integral \cite{anboot, kss} giving,
\be
z^{\D_\phi}\int_{0}^\infty \ell^{2\D_\phi-1-\ta_m}K_0(2\ell\sqrt{z})=\frac{1}{4}\G\bigg(\D_\phi-\frac{\ta_m}{2}\bigg)^2z^{\frac{\ta_m}{2}}\,.
\ee
Thus the \textit{rhs} of \eqref{bssl} becomes,
\be\label{rhs}
\frac{1}{16}\G\bigg(\D_\phi-\frac{\ta_m}{2}\bigg)^2z^{\frac{\ta_m}{2}}\sum_n \tq_{\D_\phi,n}\log v\ v^n\g_n(1-v)^{\D_\phi}F^{(d)}(2\D_\phi+2n,v)\,. 
\ee

On the \textit{lhs} of \eqref{bssl}, we will determine the coefficients of $v^\a$ as follows. To start with, we move the part $(1-v)^{\D_\phi-(d-2)/2}$ in \eqref{rhs} on the \textit{lhs} to get the $\log v$ dependent part,
\be\label{lhs1}
-\frac{P_m}{4}(1-v)^b \frac{\G(\ta_m+2\ell_m)}{\G(\ta_m/2+\ell_m)^2}\sum_{n=0}^\infty \bigg(\frac{(\ta_m/2+\ell_m)_n}{n!}\bigg)^2 v^n \log v\,,
\ee 
where  $b=\frac{\ta_m}{2}+\ell_m+\frac{d-2}{2}-\D_\phi$.
Rearranging \eqref{lhs1} as $\sum_{\a=0}^\infty B^d_\a v^\a$, where $n+k=\a$ and then summing over $k$ from $0$ to $\infty$ gives the coefficients $B^d_\a$,
\begin{align}\label{Bdm}
\begin{split}
 B^d_\a&=-\frac{4P_m\G(\ta_m+2\ell_m)}{\G(\ta_m/2+\ell_m)^2}\sum_{k=0}^\a (-1)^k\bigg(\frac{(\ta_m/2+\ell_m)_{\a-k}}{(\a-k)!}\bigg)^2\frac{b!}{k!(b-k)!}\\
&=-\frac{4P_m\G(\ta_m+2\ell_m)\G(\ta_m/2+\ell_m+\a)^2}{\G(1+\a)^2\G(\ta_m/2+\ell_m)^4}\\
&\times \ {}_3F_2\bigg(-\a,-\a,-b;1-\ell_m-\frac{\ta_m}{2}-\a,1-\ell_m-\frac{\ta_m}{2}-\a;1\bigg)\,.
\end{split}
\end{align}
The coefficient of $ v^\a\log v$ in \eqref{rhs} thus becomes,
\be
H_\a=\G\bigg(\D_\phi-\frac{\ta_m}{2}\bigg)^2\sum_{k=0}^\a 2^{2\D_\phi+2\a-2k} \tq_{\D_\phi,\a-k}\ d_{2\D_\phi+2\a-2k,k}\ \g_{\a-k}\,,
\ee
where we have put $\ta=2\D_\phi+2n$ and further $n=\a-k$ due to the fact that we have regrouped terms in $v$ with same exponent $\a$ and $d_{p,k}$ is given by \ref{ddef}. Equating $H_\a=B_\a^d$ for general $d$ dimensions\footnote{In \cite{Vos}, there was a numerical study of this recursion relation for $d=4$.}, we get following the steps in \cite{kss},
\be
\g_n=\sum_{m=0}^n \mc^d_{n,m}B_m^d\,,
\ee
where for general $d$ dimensions, the coefficients $\mc^d_{n,m}$ takes the form,
\be
\mc^d_{n,m}=\frac{(-1)^{m+n}}{8}\frac{\G(\D_\phi)^2}{(\D_\phi-d/2+1)_m{}^2}\frac{n!}{(n-m)!}\frac{(2\D_\phi+n+1-d)_m}{\G(\D_\phi-\ta_m/2)^2}\,,
\ee
where $\ta_m=d-2$ for non zero $\ell_m$ and the general $d$ dimensional coefficients $B_m^d$ for general $d$ dimensions is given in \eqref{Bdm}.

\section{Leading $n$ dependence of $\g_n$}

To find the full solution\footnote{We present the $d=6$ solution in the appendix. The $d=4$ solution can be found in \cite{kss}.} for the coefficients $\g_n$ of the anomalous dimensions in general dimensions is a bit tedious. Nonetheless we can always extract the leading $n$ dependence of the coefficients for large $n$. $\g_n$ can be written as,
\be
\g_n=\sum_{m=0}^n a_{n,m}\,,
\ee
where,
\begin{align}\label{anm}
\begin{split}
a_{n,m}=-&\frac{P_m(-1)^{m+n}\G(n+1)\G(2+2\d)\G(\D_\phi)^2\G(1+\d+m)^2\G(1+m+n-2\d+2\D_\phi)}{2\G(1+\d)^4\G(1+n-2\d+2\D_\phi)\G(n-m+1)\G(m+1)^2\G(1+m-\d+\D_\phi)^2}\\
&\times \sum_{k=0}^m \frac{(-m)_k\ ^2 (x)_k}{(-m-\d)_k\ ^2\ k!}\,,
\end{split}
\end{align}
where the summand is obtained by writing out ${}_3F_2$ in \eqref{Bdm} explicitly, $\d=d/2$ and $x=\D_\phi-2\d$. To find the leading $n$ dependence we need to extract the leading $m$ dependence inside the summation in \eqref{anm}. To do that we expand the summand around large $m$ upto $2\delta$ terms and then perform the sum over $k$ from $0$ to $m$. The large $m$ expansion takes the form,
\be
\sum_{k=0}^m\frac{(-m)_k\ ^2 (x)_k}{(-m-\d)_k\ ^2\ k!}\overset{m\gg1}{\approx}\frac{\G(2\d+1)\G(m+x+1)}{\G(m+1)\G(x+2\d+1)}+\cdots\,.
\ee
\vskip 0.5cm 
\hskip -0.5cm The terms in $\cdots$ are the subleading terms which will not affect\footnote{ This can be explicitly checked in Mathematica.} the leading order result in $n$. Thus to the leading order,

\begin{align}
\begin{split}
a_{n,m}\approx&-\frac{P_m(-1)^{m+n}\G(n+1)\G(2+2\d)\G(\D_\phi)^2\G(1+\d+m)^2\G(1+m+n-2\d+2\D_\phi)}{2\G(1+\d)^4\G(1+n-2\d+2\D_\phi)\G(n-m+1)\G(m+1)^2\G(1+m-\d+\D_\phi)^2}\\
&\times\frac{\G(2\d+1)\G(m+\D_\phi-2\d+1)}{\G(m+1)\G(\D_\phi+1)}\,.
\end{split}
\end{align}
We can now separate the parts of $a_{n,m}$ into,
\begin{align}
\begin{split}
a_{n,m}=&-\frac{P_m(-1)^{m+n}\G(2\d+1)\G(2+2\d)\G(\D_\phi)^2}{2\G(1+\d)^4\G(1+n-2\d+2\D_\phi)\G(\D_\phi+1)}\times\frac{\G(1+m+n-2\d+2\D_\phi)}{\G(1+m-\d+\D_\phi)}\\
&\times\frac{n!}{m!(n-m)!}\bigg[\frac{\G(m+\d+1)^2}{\G(m+1)^2}\frac{\G(m+\D_\phi-2\d+1)}{\G(m+\D_\phi-\d+1)}\bigg]\,.
\end{split}
\end{align}
The leading term inside the bracket is $m^\d$. Thus to the leading order in $m$, the coefficient $a_{n,m}$ is given by,
\begin{align}\label{anmd}
\begin{split}
a_{n,m}=&-\frac{P_m(-1)^{m+n}\G(2\d+1)\G(2+2\d)\G(\D_\phi)^2}{2\G(1+\d)^4\G(1+n-2\d+2\D_\phi)\G(\D_\phi+1)}\times\frac{\G(1+m+n-2\d+2\D_\phi)}{\G(1+m-\d+\D_\phi)}\\
&\times\frac{n!\ m^\d}{m!(n-m)!}\,,
\end{split}
\end{align}
\vskip 0.5 cm
\hskip -0.5cm for general $d$ dimensions. Note that to reach \eqref{anmd}, we just extracted the leading $m$ dependence without making any assumptions about $\d$ apart from $\d>0$ and $\d \in {\mathbb{R}}$. More specifically $\d$ takes integer values for even dimensions and half integer values for odd dimensions. We will now need to analyze the $m$ dependent part of $a_{n,m}$. For that note first that using the reflection formula,
\be
\G(1+m-\d+\D_\phi)\G(\d-m-\D_\phi)=(-1)^m\frac{\pi}{\sin (1-\d+\D_\phi)\pi}\,,
\ee
we can rewrite the coefficients $a_{n,m}$ as,
\begin{align}
\begin{split}
a_{n,m}=&(-1)^{n+1}\frac{\sin(\D_\phi-\d)\pi}{\pi}\frac{P_m n!\G(2\d+1)\G(2+2\d)\G(\D_\phi)^2}{2\G(1+\d)^4\G(\D_\phi+1)\G(1+n-2\d+2\D_\phi)}\\
&\frac{m^\d}{m!(n-m)!}\G(1+m+n-2\d+2\D_\phi)\G(\d-m-\D_\phi)\,.
\end{split}
\end{align}
\vskip 0.5 cm 
\hskip -0.5cm Further using the integral representation of the product of the $\G$-functions,
\be
\G(1+m+n-2\d+2\D_\phi)\G(\d-m-\D_\phi)=\int_0^\infty\int_0^\infty dx dy\ e^{-(x+y)}x^{m+n-2\d+2\D_\phi}y^{\d-m-\D_\phi-1}\,,
\ee
we pull the $m$ dependent part of $a_{n,m}$ inside the integral and perform the summation over $m$ to get,

\begin{align}\label{gin}
\begin{split}
\g_n=&(-1)^{n+1}\frac{\sin(\D_\phi-\d)\pi}{\pi}\frac{P_m n!\G(2\d+1)\G(2+2\d)\G(\D_\phi)^2}{2\G(1+\d)^4\G(\D_\phi+1)\G(1+n-2\d+2\D_\phi)}\\
&\times \int_0^\infty\int_0^\infty dx dy\ e^{-(x+y)}x^{n-2\d+2\D_\phi}y^{\d-\D_\phi-1}\sum_{m=0}^n \bigg(\frac{x}{y}\bigg)^m \frac{m^\d}{m!(n-m)!}\,.
\end{split}
\end{align}
At this stage we will need the explicit information about whether $\d$ is integer or half integer.
\vskip 1cm
\subsection{Even $d$}
For even dimensions, $\d$ is an integer. The summation over $m$ can be performed with ease now. We first put $z=\log (x/y)$. After this substitution, the summand can be written as,
\be
\sum_{m=0}^n \bigg(\frac{x}{y}\bigg)^m \frac{m^\d}{m!(n-m)!}=\sum_{m=0}^n \frac{m^\d}{m!(n-m)!}e^{m z}\,.
\ee
It is easy to see that starting from a function $f_n(z)$ and acting $\pd_z$ repeatedly on it can give back the above sum provided,
\be
f_n(z)=\sum_{m=0}^n \frac{e^{m z}}{m!(n-m)!}=\frac{(1+e^z)^n}{n!}\,.
\ee
For integer $\d$,
\be
\sum_{m=0}^n \frac{m^\d}{m!(n-m)!}e^{m z}=\pd_z^\d f_n(z)\,.
\ee
The leading order term in the derivatives of the generating function $f_n(z)$ is of the form
\be
\pd_z^\d f_n(z)=\frac{\G(n+1)}{n!\G(n+1-\d)}(1+e^z)^n(1+e^{-z})^{-\d}+\cdots\,,
\ee
where $\cdots$ represent the subleading terms in $n$. Substituting for $z$ and then into the integral representation in \eqref{gin}, we get,
\vskip 0.2cm
\begin{align}
\begin{split}
\g_n=&(-1)^{n+1}\frac{\sin(\D_\phi-\d)\pi}{\pi}\frac{P_m n!\G(2\d+1)\G(2+2\d)\G(\D_\phi)}{2\G(1+\d)^4\D_\phi\G(1+n-2\d+2\D_\phi)(n-\d)!}\\
&\times \int_0^\infty\int_0^\infty dx dy\ e^{-(x+y)}x^{n-2\d+2\D_\phi}y^{\d-\D_\phi-1}\bigg(\frac{x+y}{y}\bigg)^n\bigg(\frac{x}{x+y}\bigg)^\d\,.
\end{split}
\end{align}
The integral can be performed by a substitution of variables $x=r^2\cos^2\theta$ and $y=r^2\sin^2\theta$ and integrating over $r=0,\infty$ and $\theta=0,\frac{\pi}{2}$. This gives,

\begin{align}\label{gln}
\begin{split}
\g_n&=-\frac{\sin(\D_\phi-\d)\pi}{\pi}\frac{P_m n!\G(2\d+1)\G(2+2\d)\G(\D_\phi)}{2\G(1+\d)^4\D_\phi\G(1+n-2\d+2\D_\phi)(n-\d)!}\pi\csc(\D_\phi-\d)\pi\frac{\G(1+n-\d+2\D_\phi)}{\G(1+\D_\phi)}\\
&=-\frac{P_m\G(2\d+1)\G(2\d+2)}{2\G(1+\d)^4\D_\phi^2}\frac{\G(n+1)\G(1+n-\d+2\D_\phi)}{\G(n+1-\d)\G(1+n-2\d+2\D_\phi)}\,.
\end{split}
\end{align}
The leading term in $n$ for the last ratio in \eqref{gln} is $n^{2\d}$ giving the leading $n$ dependence of $\g_n$ for even dimensions as,
\be
\g_n=-\frac{P_m\G(2\d+1)\G(2\d+2)}{2\G(1+\d)^4\D_\phi^2}n^{2\d}\,.
\ee

\subsection{Odd $d$}
\begin{figure}
\begin{center}
\includegraphics[width=0.8\textwidth]{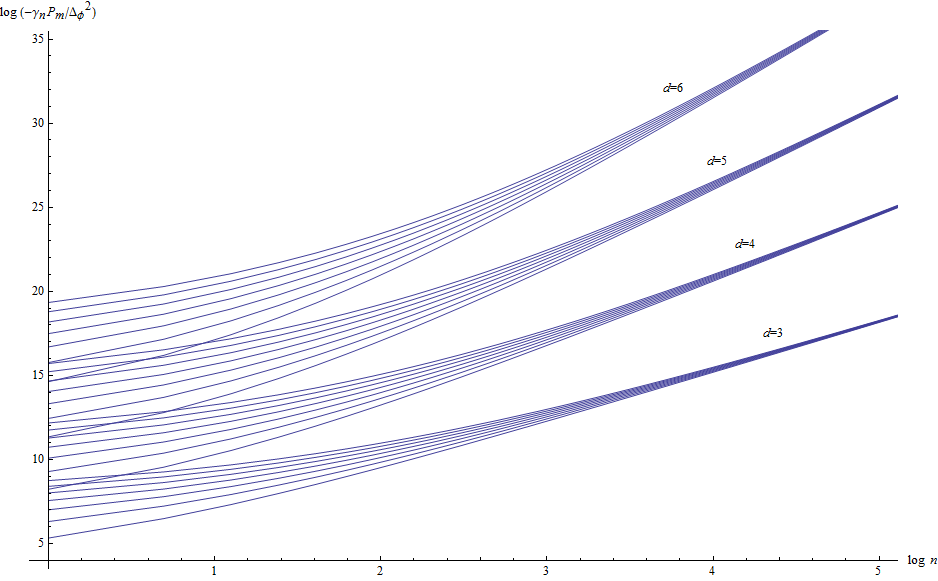}
\caption{$\log (-\g_nP_m/\D_\phi^2)$ vs. $\log n$ plot showing the dependence of $\g_n$ on $n$ for different values of $\D_\phi$ in different dimensions. For $n\gg1$, the coincidence of the graphs of different $\D_\phi$ for each $d$ indicates the universal formula given in \eqref{gl}.}
\label{fig:interceptplot}
\end{center}
\end{figure}

In odd $d$ as well we will start with \eqref{gin} but in this case the calculation is slightly different from the previous one in the sense that now $\d$ takes half integer values. We start by writing $\d=p-1/2$ where $p \in {\mathbb{Z}}$, and separate the $\sqrt{m}$ contribution by an integral representation,
\be
\frac{\sqrt{\pi}}{\sqrt{m}}=\int_{-\infty}^\infty e^{-m t^2}dt\,.
\ee
We can now write the sum in\eqref{gin} for half integer $\d$ as,
\be\label{odd}
\sum_{m=0}^n \frac{m^p \ e^{m z}}{\sqrt{m} m! (n-m)!}=\frac{1}{\sqrt{\pi}}\sum_{m=0}^n \frac{m^p}{m!(n-m)!}\int_{-\infty}^\infty e^{m(z-t^2)}dt\,.
\ee
Here $p \in {\mathbb{Z}}$. The generating function $f_n(z)$ for this case which provides with the above expression is,
\be
f_n(z)=\frac{1}{\sqrt{\pi}}\int_{-\infty}^\infty \frac{(1+e^{z-t^2})^n}{n!} dt\,,
\ee
and \eqref{odd} is,
\be
\sum_{m=0}^n \frac{m^p \ e^{m z}}{\sqrt{m} m! (n-m)!}=\pd_z^p f_n(z)\,.
\ee
The leading order term in $\pd_z^p f_n(z)$ is given by,
\be
\pd_z^p f_n(z)=\frac{1}{\sqrt{\pi}}\frac{\G(n+1)}{n!\G(n+1-p)}\int_{-\infty}^\infty dt\ \bigg\{ (1+e^{z-t^2})^{n-p}(e^{z-t^2})^{p}+\cdots\bigg\}\,,
\ee
where $\cdots$ are the subleading terms which does not affect the leading order results. We can now determine the integral by saddle point method by letting the integrand as,
\be
e^{g(t,z)}\,, \ \text{where}\ g(t,z)=(n-p)\log [1+e^{z-t^2}]+p(z-t^2)\,.
\ee
Setting $g'(t,z)=0$ we see that the saddle is located at $t=0$. Expanding around the saddle point upto second order the integrand becomes,
\be
\text{exp}[{g(t,z)}]=\text{exp}[g(0,z)+\frac{1}{2}g''(0,z)t^2+\cdots]\,,
\ee
where the terms in $\cdots$ contribute at $O(1/n)$ after carrying out the integral over $t$. Thus to the leading order,
\be
\pd_z^p f(n(z)=\frac{1}{\sqrt{\pi}}\frac{\G(n+1)}{n!\G(n+1-p)}\sqrt{\frac{2\pi}{-g''(0,z)}}e^{g(0,z)}
=\frac{\G(n+1)}{n!\G(n+1-p)}\frac{1}{\sqrt{n}}(1+e^z)^n(1+e^{-z})^{-\d}\,,
\ee
where $\d=p-1/2$. Substituting for $z$, we get the leading order in $n$ as,
\be\label{dfz}
\pd_z^p f(n(z)= \frac{n^\d}{n!} \bigg(\frac{x+y}{y}\bigg)^n\bigg(\frac{x}{x+y}\bigg)^\d\,.
\ee
We can now substitute \eqref{dfz} into \eqref{gin} and carry out the integral over $x$ and $y$ giving us back,
\be
\g_n=-\frac{P_m\G(2\d+1)\G(2\d+2)}{2\G(1+\d)^4\D_\phi^2}\frac{\G(1+n-\d+2\D_\phi)}{\G(1+n-2\d+2\D_\phi)}n^\d\,.
\ee
The large $n$ limit of the last ratio of the $\G$-functions is again $n^\d$ giving,
\be
\g_n=-\frac{P_m\G(2\d+1)\G(2\d+2)}{2\G(1+\d)^4\D_\phi^2}n^{2\d}\,.
\ee
Thus for both even and odd dimensions we get the universal result for the leading $n$ dependence for the coefficients $\g_n$,
\be\label{gl}
\g_n=-\frac{P_m\G(d+1)\G(d+2)}{2\G(1+d/2)^4\D_\phi^2}n^{d}\,.
\ee
Using $P_m=\frac{d^2\D_\phi^2}{(d-1)^2c_T}$ \cite{Komargodski},
\be\label{gammalead}
\g_n=-\frac{d^2\G(d+1)\G(d+2)}{2(d-1)^2\G(1+\frac{d}{2})^4c_T}n^d=-\frac{8(d+1)}{c_T (d-1)^2}\frac{\G(d)^2}{\G(\frac{d}{2})^4} n^d\,.
\ee
In fig.\eqref{fig:interceptplot} we show plots for $\g_n$ for various dimensions for different $\D_\phi$. At large $n$ the plots coincide which proves the universality of the leading order $n$ dependence of $\g_n$.

\section{CFT vs. Holography}
This section will focus on the matching of the findings from CFT with predictions from holography. We start by matching the CFT coefficient $P_m$ with the Newton constant $G_N$ appearing in the holographic calculations.

\subsection{Comparison with the Eikonal limit calculation}
Following \cite{Cornalba1,Cornalba2,Cornalba3} we find that in terms of the variables $h=\D_\phi+\ell+n$ and $\bh=\D_\phi+n$, the anomalous dimension $\g_{h,\bh}$ is,
\be
\g_{h,\bh}=-16G_N (h\bh)^{j-1}\Pi_\bot(h,\bh)\,,
\ee
where $\Pi(h,\bh)$ is the graviton $(j=2)$ propagator given by,
\be
\Pi_\bot(h,\bh)=\frac{1}{2\pi^{\frac{d}{2}-1}}\frac{\G(\D-1)}{\G(\D-\frac{d}{2}+1)}\bigg[\frac{(h-\bh)^2}{h\bh}\bigg]^{1-\D} \ {}_2F_1\bigg[\D-1,\frac{2\D-d+1}{2},2\D-d+1;-\frac{4h\bh}{(h-\bh)^2}\bigg]\,,
\ee
and $\D=d$ for the graviton. Thus the overall factors multiplying $\g_{h,\bh}$ are given by,
\be
\g_{h,\bh}=-\frac{8G_N}{\pi^{\frac{d}{2}-1}}\frac{\G(d-1)}{\G(\frac{d}{2}+1)} (h\bh)g(h,\bh)\,,
\ee
where $g(h,\bh)$ is the remaining function of $h$ and $\bh$. In the limit $h,\bh\rightarrow\infty$ and $h\gg\bh$,
\be\label{ghhb}
\g_{h,\bh}=-\frac{8G_N}{\pi^{\frac{d}{2}-1}}\frac{\G(d-1)}{\G(\frac{d}{2}+1)}\frac{\bh^d}{(h-\bh)^{d-2}}=-\frac{8G_N}{\pi^{\frac{d}{2}-1}}\frac{\G(d-1)}{\G(\frac{d}{2}+1)}\frac{n^d}{\ell^{\ta_m}}\,,
\ee
where $\ta_m=d-2$ for the graviton. Following \cite{Komargodski}, $G_N$ can be related to the central charge $c_T$ as,
\be\label{gnct}
G_N=\frac{d+1}{d-1}\frac{1}{2\pi c_T}\frac{\G(d+1)\pi^{d/2}}{\G(\frac{d}{2})^3}\,,
\ee
which leads to $\g_{h,\bar{h}}=-\g_n/\ell^{\tau_m}$ that matches exactly with the $\g_n$ from \eqref{gammalead}. Note that we needed $\ell\gg n\gg 1$ in the holographic calculation.

\subsection{Another gravity calculation}
In this section we will extend the calculation in \cite{kfw} to non-zero $n$. Unlike the eikonal method this approach has the advantage of considering $n=0$ as well. We leave the exact (in $n$) matching as an important and interesting open problem and address the leading order ($n \gg 1$) calculation in the present work. We list the salient features of the calculation below.
\begin{itemize}
\item{ The key idea is to write down a generic double trace operator $[\mo_1\mo_2]_{n,\ell}$ of quantum number $(n,\ell)$ formed from the descendants of operators $\mo_1$ and $\mo_2$ in the field theory, having quantum numbers $(n_1,\ell_1)$ and $(n_2,\ell_2)$ where 
\be
\ell_1+\ell_2=\ell\,, \ \ \text{and} \ \ n_1+n_2=n\,.
\ee 
}
\item{ Such an operator is given by,
\be
[\mo_1\mo_2]_{n,\ell}=\sum_{i,j}c_{ij}\pd_{\m_1}\cdots\pd_{\m_{\ell_1}}(\pd^2)^{n_1}\mo_1\pd_{\m_1}\cdots\pd_{\m_{\ell_2}}(\pd^2)^{n_2}\mo_2\,,
\ee
where $c_{ij}$ are the Wigner coefficients that depend on $n_1$, $n_2$, $\ell_1$ and $\ell_2$. For the $n=0$ case and for large $\ell$, $\ell_1=\ell_2=\ell/2$. For the $n\neq0$ it is reasonable to assume that the maxima of $c_{ij}$ will occur, in addition to the $\ell$ variable, for $n_1=n_2=n/2$. Thus $\pd^2$ will be equally distributed between the two descendents $\mo_1$ and $\mo_2$.  
}
\item{ From holography we can model the two descendants $\mo_1$ and $\mo_2$ as two uncharged scalar fields with a large relative motion with respect to each other ( corresponding to $\ell\gg1$ ) or equivalently one very massive object which behaves as an AdS-Schwarzchild\footnote{A cleaner argument may be to replace the AdS-Schwarzschild black hole with the AdS-Kerr black hole. } black hole and the other object moves around it with a large angular momentum proportional to $\ell$.  
}
\item{ The calculation of the anomalous dimension for $[\mo_1\mo_2]_{n,\ell}$ from the field theory is then equivalent to the holographic computation of the first order shift in energy of the above system of the object rotating around the black hole.
}
\end{itemize}
 The AdS-Schwarzschild black hole solution in $d+1$ dimensional bulk is given by,
\be
ds^2=-U(r)dt^2+\frac{1}{U(r)}dr^2+r^2d\Omega^2\,,
\ee
where,
\be
U(r)=1-\frac{\mu}{r^{d-2}}+\frac{r^2}{R_{AdS}^2}\,.
\ee
The mass of the black hole is given by,
\be\label{bhmass}
M=\frac{(d-1)\Omega_{d-1}\mu}{16\pi G_N}\,.
\ee
The first order shift in the energy is given by\footnote{ The normalization outside should be $1/2$ and not $1/4$ as used in \cite{kfw}. We thank Jared Kaplan for confirming this.},
\begin{align}\label{dE}
\begin{split}
\d E_{orb}^d&=\la n,\ell_{orb}|\d H|n,\ell_{orb}\ra\\
&=-\frac{\mu}{2}\int r^{d-1}dr d^{d-1}\Omega\la n,\ell_{orb}|\bigg(\frac{r^{2-d}}{(1+r^2)^2}(\pd_t\phi)^2+r^{2-d}(\pd_r\phi)^2\bigg)|n,\ell_{orb}\ra\,.
\end{split}
\end{align}
The label `\textit{orb}' implies that currently we are considering one of the masses of the binary system. We can also add higher derivative corrections coming from the $\a'$ corrections to the metric. This is one of the advantages of doing the anomalous dimension calculation in this approach. It will make transparent the fact that the $\a'$ corrections will not affect the leading result. The metric will be modified\footnote{There will also be an overall $\alpha'$ dependent factor which can be absorbed into $c_T$ \cite{ss}.} by adding corrections to the factor $r^{2-d}(1+c_h\a'^h r^{-2h})$ where $h$ is the order of correction in $\a'$.  The wavefunction of the descendant state derived from the primary is given by,
\be\label{wvfn}
\psi_{n,\ell J}(t,\r,\Omega)=\frac{1}{N_{\Delta_\phi n\ell}}e^{-iE_{n,\ell}t}Y_{\ell,J}(\Omega)\bigg[\sin^{\ell} \r\cos^{\Delta_\phi}\r\ {}_2F_1\bigg(-n,\Delta_\phi+\ell+n,\ell+\frac{d}{2},\sin^2\r\bigg)\bigg]\,,
\ee 
where $E_{n,\ell}=\Delta_\phi+2n+\ell$ and,
\be
N_{\Delta_\phi n\ell}=(-1)^n\sqrt{\frac{n!\G^2(\ell+\frac{d}{2})\G(\Delta_\phi+n-\frac{d-2}{2})}{\G(n+\ell+\frac{d}{2})\G(\Delta_\phi+n+\ell)}}\,.
\ee
Using the transformation $\tan\r=r$ we can write the scalar operator as,
\begin{align}\label{poly0}
\begin{split}
\psi_{n,\ell_{orb} J}(t,r,\Omega)&=\frac{1}{N_{\Delta_\phi n\ell_{orb}}}e^{-iE_{n,\ell_{orb}}t}Y_{\ell_{orb},J}(\Omega)\bigg[\frac{r^{\ell_{orb}}}{(1+r^2)^{\frac{\Delta_\phi+\ell_{orb}}{2}}}\sum_{k=0}^n\frac{(-n)_k(\Delta_\phi+\ell_{orb}+n)_k}{(\ell_{orb}+\frac{d}{2})_k\ k!}\bigg(\frac{r^2}{1+r^2}\bigg)^k\bigg]\\&=\sum_{k=0}^n \psi^{orb}_k(t,r,\Omega)\,,
\end{split}
\end{align}
\vskip -0.6cm
\hskip -0.5cm where,
\be\label{psi}
\psi^{orb}_k(t,r,\Omega)=\frac{1}{N_{\Delta_\phi n\ell_{orb}}}e^{-iE_{n,\ell_{orb}}t}Y_{\ell_{orb},J}(\Omega)\bigg[\frac{r^{\ell_{orb}}}{(1+r^2)^{\frac{\Delta_\phi+\ell_{orb}}{2}}}\frac{(-n)_k(\Delta_\phi+\ell_{orb}+n)_k}{(\ell_{orb}+\frac{d}{2})_k\ k!}\bigg(\frac{r^2}{1+r^2}\bigg)^k\bigg]\,.
\ee
Putting \eqref{psi} in \eqref{dE} and carrying out the other integrals we are left with just the radial part of the integral,
\be
\d E_{orb}^d=-\mu\int r(1+c_h\a'^h r^{-2h})dr\ \bigg[\sum_{k,\a=0}^n\bigg(\frac{E_{n,\ell_{orb}}^2}{(1+r^2)^2} \psi^{orb}_k(r)\psi^{orb}_\a(r)+\pd_r\psi^{orb}_k(r)\pd_r\psi^{orb}_\a(r)\bigg)\bigg]={\mathcal{I}}_1+{\mathcal{I}}_2\,,
\ee
where ${\mathcal{I}}_1$ and ${\mathcal{I}}_2$ are the contributions from the first and the second parts of the above integral. 
The leading $\ell_{orb}$ dependence comes from the first part of the integral which is also true for $n\neq0$ case and for any general $d$ dimensions is given by $(1/\ell_{orb})^{(d-2)/2}$. Thus we can just concentrate on the first part of the integral for the leading spin dominance of the energy shifts. Thus the integral ${\mathcal{I}}_1$ can be written as,
\begin{align}
{\mathcal{I}}_1=-\frac{\mu}{N_{\Delta_\phi n\ell_{orb}}^2}&\sum_{k,\a=0}^n \frac{(-n)_k(-n)_\a(\Delta_\phi+n+\ell_{orb})_k(\Delta_\phi+n+\ell_{orb})_\a}{(\ell_{orb}+\frac{d}{2})_k(\ell_{orb}+\frac{d}{2})_\a\ k!\a!}\nonumber\\
&\times\int_0^\infty r(1+c_h\a'^hr^{-2h}) dr \frac{r^{2\ell_{orb}+2k+2\a}}{(1+r^2)^{2+\Delta_\phi+\ell_{orb}+k+\a}}\,.
\end{align}
\vskip -0.4cm
\hskip -0.5cm The $r$ integral gives,
\begin{align}
&\int_0^\infty r(1+c_h \a'^hr^{-2h})dr \frac{r^{2\ell_{orb}+2k+2\a}}{(1+r^2)^{2+\Delta_\phi+\ell_{orb}+k+\a}}\nonumber\\
&=\frac{\G(1+\Delta_\phi)\G(1+\ell_{orb}+k+\a)+c_h\a'^h \G(1+h+\Delta_\phi)\G(1+\ell_{orb}+k+\a-h)}{2\G(2+\Delta_\phi+\ell_{orb}+k+\a)} \,.
\end{align}
Hence the integral ${\mathcal{I}}_1$ becomes,
\begin{align}
\begin{split}
{\mathcal{I}}_1=&-\frac{\mu}{2N_{\Delta_\phi n\ell_{orb}}^2}\sum_{k,\a=0}^n  \frac{(-n)_k(-n)_\a(\Delta_\phi+n+\ell_{orb})_k(\Delta_\phi+n+\ell_{orb})_\a}{(\ell_{orb}+\frac{d}{2})_k(\ell_{orb}+\frac{d}{2})_\a\ k!\a!}\\
&\times\frac{\G(1+\ell_{orb}+k+\a)\G(1+\Delta_\phi)+c_h\a'^h\G(1+h+\Delta_\phi)\G(1+\ell_{orb}+k+\a-h)}{\G(2+\Delta_\phi+\ell_{orb}+k+\a)}\,.
\end{split}
\end{align}
Using the reflection formula for the $\G$-functions we can write,
\be
(-n)_k(-n)_\a=(-1)^{k+\a}\frac{\G(n+1)^2}{\G(n+1-k)\G(n+1-\a)}\,.
\ee
Putting in the normalization and performing the first sum over $\a$ we get,
\begin{align}\label{adsd}
\begin{split}
{\mathcal{I}}_1=&-\frac{\mu (\ell_{orb}+2n)^2\G(\ell_{orb}+\frac{d}{2}+n)}{2\G(\ell_{orb}+\frac{d}{2})\G(1-\frac{d}{2}+n+\Delta_\phi)}\sum_{k=0}^n\frac{(-1)^k\G(k+\ell_{orb}+n+\Delta_\phi)}{\G(\ell_{orb}+\frac{d}{2}+k)\G(n+1-k)\G(2+k+\ell_{orb}+\Delta_\phi)\G(k+1)}\\
&\times\ \bigg[\G(1+\ell_{orb}+k)\G(1+\Delta_\phi){}_3F_2\bigg(-n,k+\ell_{orb}+1,\ell_{orb}+n+\Delta_\phi;\ell_{orb}+\frac{d}{2},2+k+\ell_{orb}+\Delta_\phi;1\bigg)\\
&+c_h\a'^h\G(1+\ell_{orb}+k-h)\G(1+\Delta_\phi+h)\\
&\times{}_3F_2\bigg(-n,k+\ell_{orb}+1-h,\ell_{orb}+n+\Delta_\phi;\ell_{orb}+\frac{d}{2},2+k+\ell_{orb}+\Delta_\phi;1\bigg)\bigg]\,.
\end{split}
\end{align}
To the leading order in $\ell_{orb}$ (after the exapnsion in large $\ell_{orb}$) this is the expression for $\d E_{orb}^d$ or equivalently $\g_{n,\ell_{orb}}$ from the CFT for general $d$ dimensions.  Since the $\ell_{orb}$ dependence does not rely on $n$ dependence we can use the $n=0$ to show the $\ell_{orb}$ dependence from the $\a'$ corrections. Thus in \eqref{adsd} we put $n=0$ to get, 
\be
{\mathcal{I}}_1=-\frac{\m}{2\G(1-\frac{d}{2}+\D_\phi)}\bigg(\frac{1}{\ell_{orb}}\bigg)^{\frac{d-2}{2}}\bigg[1+c_h\a'^h\bigg(\frac{1}{\ell_{orb}}\bigg)^h\bigg]\,.
\ee
In IIB string theory $\a'$ corrections to the metric start $h=3$ due to the well known $R^4$ term. Thus the $\a'$ corrections contribute at a much higher order, $O(1/\ell_{orb}^3)$, in $1/\ell_{orb}$ and hence does not affect the leading order result.  In the rest of the work, we will consider only the leading order result in $\ell_{orb}$ (set $c_h=0$) for general $d$. 

For even $d$ we can calculate ${\mathcal{I}}_1$ for $n=0,1,2\cdots$ and infer a general $n$ dependent polynomial. Using this polynomial, the leading $n$ dependence of the energy shifts become,
\be\label{poly}
\d E_{orb}^d=-\m\bigg(\frac{1}{\ell_{orb}}\bigg)^{\frac{d-2}{2}}\bigg(\frac{\G(d)}{\G(\frac{d}{2})\G(\frac{d}{2}+1)}n^{d/2}+\cdots\bigg)\,,
\ee
where $\cdots$ are terms subleading in $n$. For odd $d$ the polynomial realization is less obvious due to factors of $\G(\D_\phi+\frac{1}{2})$ etc. nevertheless one can show ( see fig.2) at least numerically that the leading order $n$ dependence for odd $d$ should also be the same as \eqref{poly}. Hence the above form in \eqref{poly} is true for general $d$ dimensions. 

\begin{figure}[!htpb]
\begin{subfigure}{.3\textwidth}
  \centering
  \includegraphics[width=\textwidth]{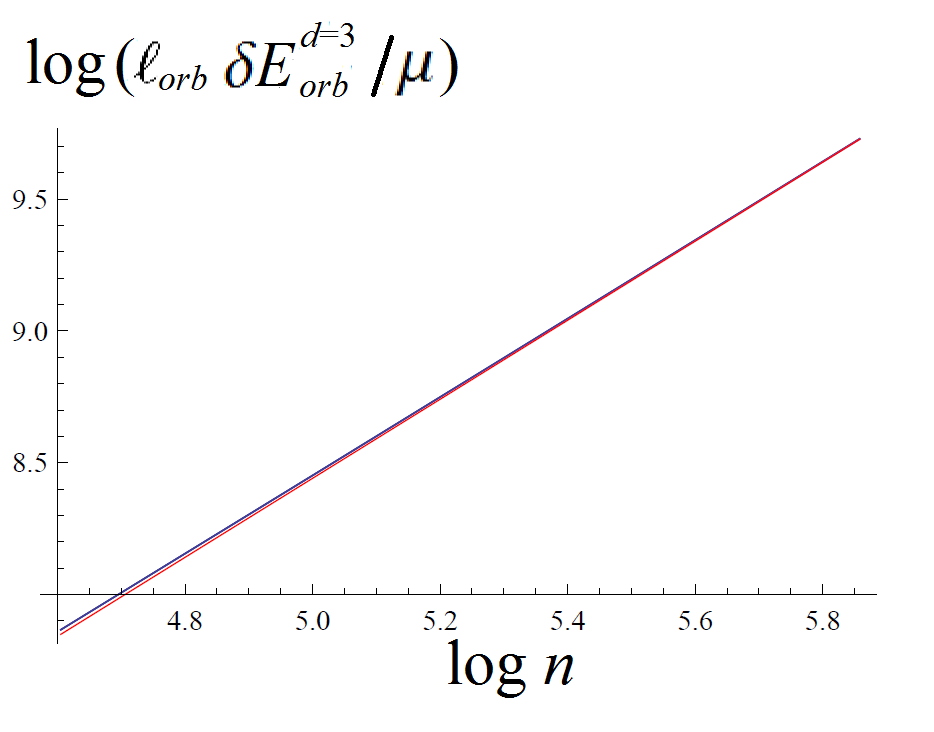}
  \caption{$d$=3}
  \label{fig:sub1}
\end{subfigure}\hfill%
\centering
\begin{subfigure}{0.3\textwidth}
  \centering
  \includegraphics[width=\textwidth]{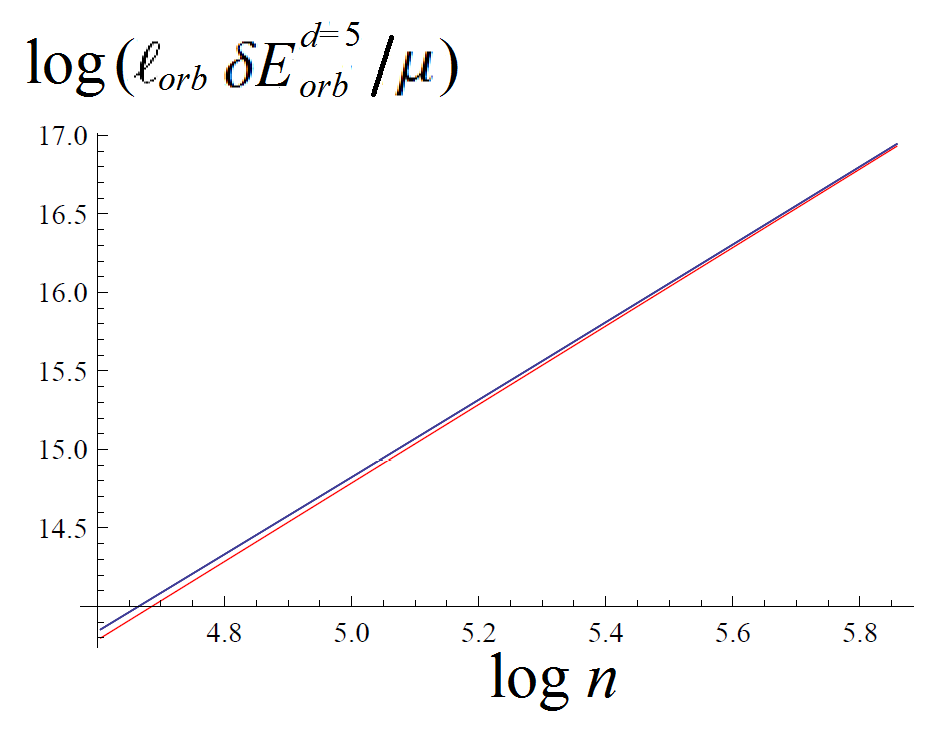}
  \caption{$d$=5}
  \label{fig:sub2}
\end{subfigure}\hfill%
\begin{subfigure}{0.3\textwidth}
  \centering
  \includegraphics[width=\textwidth]{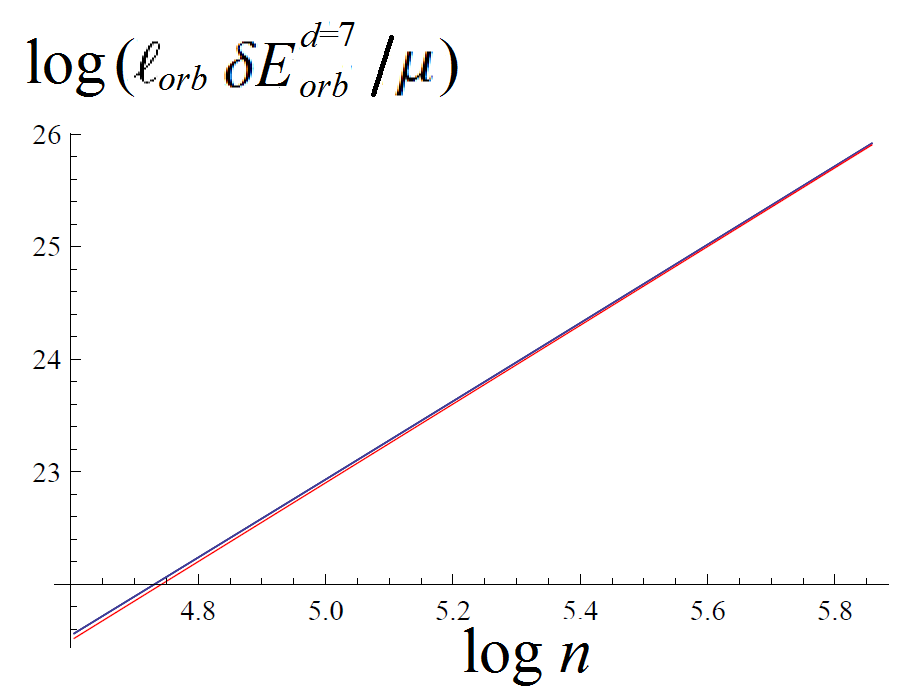}
  \caption{$d$=7}\label{n8n9}
  \label{fig:sub3}
\end{subfigure}\hfill
\caption{The blue lines are $\log (\ell_{orb}^{d/2 -1}\delta E^d_{orb}/\mu) $ vs $\log n$ for large values of $\ell_{orb}$ and the red lines are the log-log plot for $rhs$ of \eqref{poly}, in odd $d$. The matching of both lines at large $n$ is the numerical proof of \eqref{poly} in odd $d$.}
\label{otherns}
\end{figure}

Note that we are retrieving only half of the $n$ dependence from \eqref{poly}. The other half will come from the definitions of $\ell_{orb}$, its relation with $\ell$ and $\m$ which is related to the black hole mass \textit{via} \eqref{bhmass}. The relation between $\ell_{orb}$ and $\ell$ can be derived following \cite{kfw}. For the orbit state of an object rotating around a static object, $\ell_{orb}$ is related to the geodesic length $\kappa$ that maximizes the norm of the wave function (\ref{wvfn}). We can approximate the hypergeometric function appearing in (\ref{wvfn}) by taking the large $\ell$ approximation in each term of the sum in (\ref{poly0}), 
\begin{align}
{}_2F_1\bigg(-n,\Delta_\phi+\ell_{orb}+n,\ell_{orb}+\frac{d}{2},\sin^2\r\bigg)\bigg] \ &=\sum_{k=0}^n\frac{(-n)_k(\Delta_\phi+\ell_{orb}+n)_k}{(\ell_{orb}+\frac{d}{2})_k\ k!}(\sin^2 \rho)^k\nonumber\\& \overset{\ell_{orb}\to\infty}{\approx} \ \sum_{k=0}^n \frac{(-n)_k \ell_{orb}^k}{\ell_{orb}^k k!} \sin^{2k} \rho \  = \  \cos^{2n}\rho\,.
\end{align}
Using this approximation in (\ref{wvfn}) we find the maxima of $\psi_{n,\ell_{orb} J}^2$, which occurs at,
\be
\rho=\tan^{-1}\sqrt{\frac{\ell_{orb}}{2n+\D_\phi}}\,.
\ee
Now using the relation, $\sinh \k = \tan \rho$, we find for large $\ell_{orb}\footnote{Here there is a factor of 2 mismatch with the corresponding result given in \cite{kfw}. We thank Jared Kaplan for pointing out that this was a typo in \cite{kfw}. }$,
\be\label{1stat}
\kappa=\frac{1}{2} \log \left(\frac{4 \ell_{orb}}{\Delta_\phi+n}\right)
\ee
But our case is that of a double trace primary operator where none of the two objects is static.
However in the semi-classical limit the energy shift of the former case is same as that of the primary if the geodesic distance between the two objects is same in both cases. This gives, $\kappa = \kappa_1 +\kappa_2$, where
\be\label{2stat}
\kappa_1 \  (=\kappa_2) \ =\frac{1}{2} \log \frac{4\ell_1}{\Delta_\phi+n}\,.
\ee
Since we have a composite operator which is like two particles rotating around each other, the conformal dimension of each descendant state gets the maximum contribution when $\ell_1 \approx \ell_2 \approx \ell/2$ and $n_1\approx n_2 \approx n/2$ for large $\ell$ and $n$. This is why we have $\ell/(\D_\phi+n)$ inside log in the above equation. From \eqref{1stat} and \eqref{2stat} we get, $\ell_{orb} \approx \ell^2/n$ for large $n$. So, the other factor sitting in front of (\ref{poly}) is given by,
\be
\frac{\mu}{{\ell_{orb}}^{(d-2)/2}}=\frac{2 G_N M \pi ^{(d-2)/2} \Gamma \left(\frac{d}{2}\right)}{d-1} \ \frac{n^{d/2}}{\ell^{d-2}}
\ee
Here we have used the equation for black hole mass \eqref{bhmass}. Since $M$ relates to the case of one massive static object at the centre, we will put $M=\Delta_\phi+n \approx n$. Using equation (\ref{gnct}) we get the anomalous dimension to be,
\begin{align}
\gamma(n,\ell)=-\frac{8(d+1)}{c_T (d-1)^2}\frac{\G(d)^2}{\G(\frac{d}{2})^4}\frac{n^d}{\ell^{d-2}}\,.
\end{align}
which is the same as that found from CFT.

\section{Discussions}
We conclude by listing some open questions and interesting future problems:
\begin{itemize}
\item{ One obvious extension of our analysis is to find the full $n$ dependent expression for arbitrary dimensions. We know that closed form expressions exist for even dimensions (see the appendix for the $d=6$ result) but it will be interesting to see the analogous expressions for odd dimensions. It should also be possible to repeat our analysis for general twists $\tau_m$. We restricted our attention to the case where this was the stress tensor.}
\item{The $n\gg\ell$ case for arbitrary dimensions which we expect to work out in a similar manner following \cite{kss}. In this case we needed $\ell/n$ bigger than some quantity which on the holographic side was identified with a gap--this was similar to the discussion in \cite{Camanho:2014apa}. This means that for operators with $\ell/n$ smaller than the gap, the result would be sensitive to $\alpha'$ corrections. It could be possible that the anomalous dimensions of these operators would not be negative indicating a problem with causality in the bulk. It will be interesting to extend our holographic calculations to $n\gg \ell\gg 1$.}
\item{One should compare our results with those from the numerical bootstrap methods eg. in \cite{Showk}.}
\item{One could try to see if there exists a simple argument for the leading $n$ dependence following the lines of \cite{Alday2}.}
\item{It will be interesting to match the CFT and holographic calculations to all orders in $n$ at least for a class of theories (e.g. $\mathcal{N}=4$ results in \cite{Alday}). }

\end{itemize}

\section*{Acknowledgments} We thank Fernando Alday, Jared Kaplan, Shiraz Minwalla and Hugh Osborn for discussions and correspondence.  We thank Fernando Alday and Jared Kaplan for useful comments on the draft. AS thanks Swansea University and Cambridge University for hospitality during the course of this work. AS acknowledges support from a Ramanujan fellowship, Govt. of India.

\appendix

\section{Exact $n$ dependence in $d=6$}
Here we will use the general $d$ expressions discussed in section \ref{anomdimgen}. For general $d$, $\g_n$ is given by,
\begin{align}\label{d6main}
\g_n=\sum_{m=0}^n&\frac{\left(4P_m\right)(-1)^{m+n} n \Gamma (2+d) \Gamma ^2\left(1+\frac{d}{2}+m\right) \Gamma ^2\left(\Delta _{\phi }\right) (1-n)_{m-1} (d-m-n-2 \Delta_ \phi )_m}{8\Gamma ^2\left(\Delta _{\phi }-\frac{d-2}{2}\right) \Gamma ^4\left(1+\frac{d}{2}\right) \Gamma ^2(1+m) \left(1-\frac{d}{2}+\Delta _{\phi }\right)_m}\nonumber\\ &\times {}_3F_2\left[\left(
\begin{array}{c}
 -m,-m,-d+\Delta_\phi  \\
 -\frac{d}{2}-m,-\frac{d}{2}-m
\end{array}
\right),1\right] \,.
\end{align}
In $d=6$ the hypergeometric function can be written as,
\begin{align}\label{d6hyp}
& {}_3F_2\left[\left(
\begin{array}{c}
 -m,-m,-d+\Delta_\phi  \\
 -\frac{d}{2}-m,-\frac{d}{2}-m
\end{array}
\right),1\right] =\sum_{k=0}^m \frac{(1-k+m)^2 (2-k+m)^2 (3-k+m)^2 \Gamma\left(-6+k+\Delta _{\phi }\right)}{(1+m)^2 (2+m)^2 (3+m)^2 k! \Gamma\left(-6+\Delta _{\phi }\right)}\nonumber\\&
=\frac{36 \Gamma \left(-2+m+\Delta _{\phi }\right) \left(-2+2 m+\Delta _{\phi }\right) \left(10 (-2+m) m+\Delta _{\phi } \left(-1+10 m+\Delta _{\phi }\right)\right)}{(1+m) (2+m) (3+m) \Gamma (4+m) \Gamma \left(1+\Delta _{\phi }\right)}\,.
\end{align}
\vskip 0.5cm
\hskip -0.5cm To see how to get the above result, we will use the tricks introduced in the appendix of \cite{kss}. Note that we can write,
\begin{align}
(1-k+m)^2 (2-k+m)^2 (3-k+m)^2= A k(k-1)(k-2)(k-3)(k-4)(k-5)\nonumber\\ +B k(k-1)(k-2)(k-3)(k-4)+C k(k-1)(k-2)(k-3)\nonumber\\+D k(k-1)(k-2)+E k(k-1) +F k+G\,.
\end{align}
where,
\vskip -1cm
\begin{align}
&A=1, \  B=3-6 m, \ C=3+15 m^2, \ D=-2 (1+2 m) (3+5 m (1+m)),\nonumber\\&E=3 (1+m)^2 (6+5 m (2+m)), F=-3 (1+m)^2 (2+m)^2 (3+2 m),\nonumber\\&G=(1+m)^2 (2+m)^2 (3+m)^2\,.
\end{align}
So the summation in (\ref{d6hyp}) breaks up into seven different sums of the general form shown below,
\be
\sum_{k=0}^m k(k-1)\cdots (k-i+1)\frac{\G(x+k)}{k!\G(x)}=\frac{\Gamma (m+x+1)}{(x+i)\Gamma (m-i+1)\Gamma (x)}\,.
\ee
Once again the above result was derived in \cite{kss}. With this and summing up the respective terms, the hypergeometric funtion simplifies to the form shown in the second line of (\ref{d6hyp}).

The main equation \eqref{d6main} in $d=6$ has the form,
\begin{align}
\g_n=\sum_{m=0}^n &\frac{70 (-1)^{n+1} P_m n!\text{  }\Gamma \left(-5+m+n+2 \Delta _{\phi }\right)\Gamma \left(3-m-\Delta _{\phi }\right) \Gamma \left(\Delta _{\phi }\right)\sin \left(\Delta _{\phi } \pi \right)}{\Delta _{\phi } m! (n-m)!\Gamma \left(-5+n+2 \Delta _{\phi }\right)\pi }\nonumber\\ &\times \left(-2+2 m+\Delta _{\phi }\right) \left(10 m^2+10 m \left(-2+\Delta _{\phi }\right)+\left(-1+\Delta _{\phi }\right) \Delta _{\phi }\right)\,.
\end{align}
\begin{figure}[ht]
\begin{center}
\includegraphics[width=0.65\textwidth]{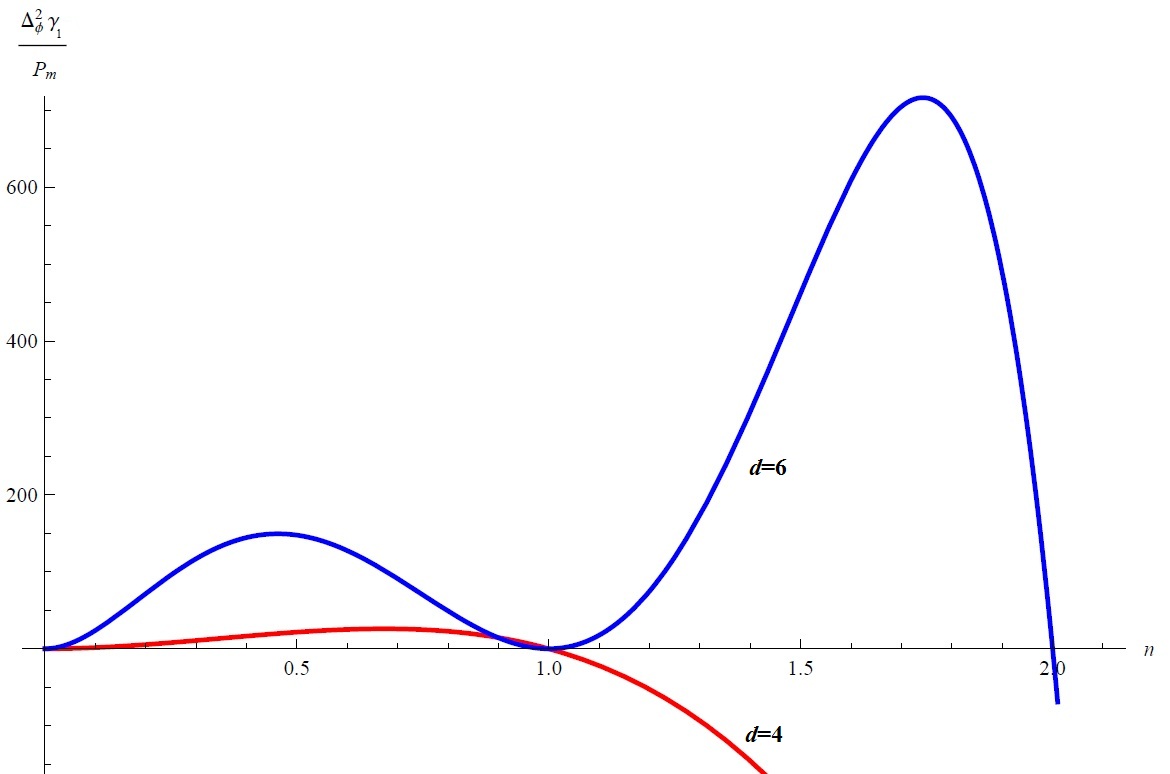}
\caption{$\g_{n=1}(\D_{\phi}^2/P_m)$ vs $\D_{\phi}$ plot in $d=4$ (red curve) and $d=6$ (blue curve), showing that the anomalous dimension may take positive values if unitarity is violated. Unitarity demands $\D_{\phi}\ge (d-2)/2$.}
\label{fig:unitarityplot}
\end{center}
\end{figure}
To get the above, we have used the reflection formula of gamma functions, $\Gamma(1-z)\Gamma(z)=\pi/\sin(\pi z)$.  To do the sum we write the gamma functions in the numerator as,
\be\label{xyint}
\Gamma (3-m-\Delta \phi ) \Gamma (-5+m+n+2 \Delta \phi )=\int _0^{\infty }\int _0^{\infty }dx dy \  e^{-(x+y)}x^{-6+m+n+2 \Delta_\phi }y^{2-m-\Delta_\phi }\,.
\ee
Let us write $f(m)=\left(-2+2 m+\Delta _{\phi }\right) \left(10 m^2+10 m \left(-2+\Delta _{\phi }\right)+\left(-1+\Delta _{\phi }\right) \Delta _{\phi }\right)$. Then the sum over $m$ becomes,
\begin{align}
&\sum _{m=0}^n \frac{(x/y)^mf(m)}{m! (n-m)! }=\frac{(x+y)^n }{y^n(x+y)^3 n!}\left(20 n^3 x^3+(x+y)^3 (-2+\Delta \phi ) (-1+\Delta \phi ) \Delta \phi \right.\nonumber\\ &\left.+30 n^2 x^2 (x (-2+\Delta \phi )+y \Delta \phi )+2 n x \left(20 x^2-3 (x+y) (7 x+2 y) \Delta \phi +6 (x+y)^2 \Delta \phi ^2\right)\right)\,.
\end{align}
The final result can now be obtained by integrating over $x$ and $y$. To do this, we change the variables to $x=r^2\cos^2\theta$ and $y=r^2\sin^2\theta$. Then we get,
\begin{align}
\g_n=&-\frac{70 (-1)^n P_m n!\text{  }\Gamma \left(\Delta _{\phi }\right)\sin \left(\Delta _{\phi } \pi \right)}{\Delta _{\phi } m! (n-m)!\Gamma \left(-5+n+2 \Delta _{\phi }\right)\pi }\int _0^{\infty }\int _0^{\infty }dx \ dy e^{-(x+y)}x^{-6+n+2 \Delta \phi }y^{2-\Delta \phi }\sum _{m=0}^n\frac{(x/y)^mf(m)}{m! (n-m)! }\nonumber\\=&-\frac{70}{\Delta_\phi ^2} P_m\left(20n^6+60 \left(-5+2 \Delta _{\phi }\right)n^5+10 \left(170-132 \Delta _{\phi }+27 \Delta _{\phi }^2\right)n^4\right.\nonumber\\&+20 \left(-5+2 \Delta _{\phi }\right) \left(45-32 \Delta _{\phi }+7 \Delta _{\phi }^2\right)n^3+2 \left(2740-3780 \Delta _{\phi }+2121 \Delta _{\phi }^2-582 \Delta _{\phi }^3+66 \Delta _{\phi }^4\right)n^2\nonumber\\&\left.+4 \left(-5+2 \Delta _{\phi }\right) \left(120-140 \Delta _{\phi }+73 \Delta _{\phi }^2-21 \Delta _{\phi }^3+3 \Delta _{\phi }^4\right)n+\left(-2+\Delta _{\phi }\right){}^2 \left(-1+\Delta _{\phi }\right){}^2 \Delta _{\phi }^2\right)\,.
\end{align}

This is the result for $\g_n$ in $d=6$. Note that we simply extended the tricks used for $d=4$ in the appendix of \cite{kss}. The same can be done to evaluate $\g_n$ for any even $d>2$.

The figure \ref{fig:unitarityplot} shows the range of $\D_\phi$, in $d=4$ and  $6$ for which $\g_{n=1}$ takes positive values. Interestingly, for both the cases positive values of $\g_1$ are over those values of $\D_\phi$ which violate the respective unitarity bounds. In fact this behaviour of the anomalous dimensions is true in any dimension. We found that for $n\geq d-1$, $\g_n$ was negative for any $\D_\phi>0$.

\end{document}